# Design and Evaluation of Distributed Networked Control for a Dual-Machine Power System

Babak Tavassoli

**Abstract:** Oscillations between swing modes of electric machines is an important limitation in achieving a high level of transient performance and reliability in power grids. Based on the new advances in measurement and transmission of wide-area information, this work proposes a distributed networked control scheme by considering the communication delays. The results are applied to reduce the inter-area swing oscillations in a power grid. In comparison with the previous works, we provide a more realistic modeling of the resulting networked control system with data sampling and delays. The exactness of the proposed modeling allows for precise evaluation and comparison between the distributed and decentralized schema. A symmetric a dual machine power system is highly oscillatory and we focus on this case to evaluate the ability of the proposed control design in dampening of the oscillations. The design can be done either based on optimization of a quadratic cost function or a disturbance attenuation level.

## I. INTRODUCTION

Contemporarily, in many of the systems applications it is possible to have access to remotely measured information relatively in real-time. This has been made possible by the advances in the networked communication technologies that will continue to provide higher levels of access to remote data. This kind of information access directly influences the control systems in many of the applications by moving the traditionally local controllers toward controllers that use the remote data to achieve higher capabilities. Inspired by the underlying network of interconnections, such a global system is sometimes referred to as a networked control system (NCS) [1, 2, 3].

Power grids are one of the largest types of systems that have been invented by the human. For performance and transient stability analysis purposes, the power network needs to be studied as a whole due to the highly interactive nature of its components. Another issue is the growth of a power network along with time which moves it away from its properly designed initial structure and causes degradation of performance, stability and reliability. Based on these facts, it is expected that a networked control scheme is able to achieve a superior performance compared to the traditionally decentralized controllers that use only the locally measured information. Development of data networking protocols for power systems [4, 5, 6] in parallel with wide area measurement (WAM) technologies [7, 8, 9] that have been initially intended for monitoring purposes can serve the power system controllers. Also, control of the recently introduced flexible AC transmission systems (FACTS) basically requires remote (wide area or global) feedback signals [10]. Early results in the direction of designing networked controllers for power grids include [11, 12] in which centralized control based on WAM is designed without considering the limitations of data sampling/transmission rate and communication delays. However, these phenomena have crucial effects on the control performance and must not be neglected. In [13, 14], delays are taken into consideration by assuming that the delays of different links have the same value and centralized control is designed using the Smith predictor method. Other works include application of Lyapunov based techniques for time delay systems [15, 16, 17, 18] that can guarantee the stability and performance with conservativeness.

In this paper we turn to the question that "To what extent the control performance can be improved by applying a networked control scheme?". To the best of our knowledge none of the previous works provide an exact and clear answer to this question. We will aim to study the ability of the networked controller to reduce two different types of quantitative measure including:

- A cost functional of the system error variables.
- A disturbance attenuation criterion.

The Lyapunov based techniques are not suitable for our purpose due to their conservative nature. To calculate the above performance measures, we follow an exact modeling and analysis procedure. The effects of sampling and arbitrary delays are captured in a time-discretization framework. The selected modeling approach allows for calculation of the above measures by solving some algebraic Riccati equations (instead of calculation bounds on the value of measures based on sufficient conditions in the form of matrix inequalities). By considering the inter-area oscillations as a form of interactive resonance between system components, the most oscillatory case is when the resonance frequency of the components coincide. Therefore, we consider two similar synchronous generators connected to a symmetrical network. Also, we will use the distributed control structure instead of the centralized control in the mentioned works. Each machine has a local stabilizing controller which also receives a remote command signal that improves the system performance. It is mentioned that a distributed system is less prone to failures.

The organization of the paper is as follows. In the next section the continuous-time modeling of the multi-machine power grid is considered. In section three, a framework for sampling with delays and some necessary tools for the analysis are provided. In section four, the distributed network control scheme is proposed which is applied to the symmetric power grid in section five to study the effects of the delays on the performance measures. Conclusions are made at the end.

*Notation*: The set of positive integers, non-negative integers, real numbers, and non-negative real numbers are denoted by $\mathbb{N}$, $\mathbb{Z}_0^+$, $\mathbb{R}$, $\mathbb{R}_0^+$, respectively. The dependency of signals on the continuous time *t* is omitted when there is no ambiguity. An identity matrix of dimension $n \times n$ is denoted by $I_n$, a zero matrix of dimension $n \times m$ is denoted by $0_{n \times m}$ or $0_n$ if $n = m$. The indices may be dropped if the dimensions can be determined from the context.

## II. MULTI-MACHINE POWER SYSTEM

In this section, the overall small signal modeling of a power system is considered which is composed of multiple synchronous generators, loads, transmission lines and other components (transformers, etc.). We consider each generator as a subsystem of our model and integrate the rest of power system into a single additional subsystem which will be denoted as the network. Thus, if the number of generators is $m \in \mathbb{N}$, then the number of subsystems in the model becomes $m+1$. Since we start the modeling process from the basic equations of the generators, it was found easier to work with physical units instead of the per unit systems to avoid an increased number of variables and relations. The reference for the contents of this section is [19].

### A. synchronous generators

The physical behavior of a synchronous generator is described by its basic equations. But, the traditional approach is to model the generator dynamics based on its observed behavior to facilitate the identification of its parameters [19]. In here, we use the basic equations to build our model relying on the new advanced system identification methods. The swing equations of a generator are

$$\dot{\delta}_i = \omega_i - \omega_0 \tag{1.1}$$

$$J_i \dot{\omega}_i = T_{m,i} - T_{e,i} - B_i \omega_i \tag{1.2}$$

where the index *i* denotes the *i*th generator, $\omega_i$ is the rotor speed, $\delta_i$ is the angular position of the rotor with respect to a frame rotating with the constant synchronous speed $\omega_0$ for the whole power grid, $T_{m,i}$ is the mechanical torque applied to the generator which will be assumed to be constant (since its slow dynamics do not contribute to the oscillations), $T_{e,i}$ is the electrical torque produced by the generator, $J_i$ is the moment of inertia of the rotor and $B_i$ is a friction coefficient.

The stator equations are usually written in a rotating *dq* frame. We set the speed of the *dq* frame to the synchronous speed $\omega_0$ to be able to compose the generator model with the network. In the frequency range for the swing oscillations the amortisseur windings can be effectively neglected and the flux and voltage equations of the rotor are written as

$$\psi_{f,i} = L_{f,i} i_{f,i} - \frac{3}{2} L_{af,i} [\cos(\delta_i) \quad \sin(\delta_i)] \begin{bmatrix} i_{d,i} \\ i_{q,i} \end{bmatrix} \tag{2.1}$$

$$e_{f,i} = \dot{\psi}_{f,i} + R_{f,i} i_{f,i} \tag{2.2}$$

where $\psi_{fd,i}$ is the flux of the rotor field winding, $i_{f,i}$ is the current of the field winding, $e_{f,i}$ is the voltage applied to the field winding, $i_{d,i}$ and $i_{q,i}$ are the *d* and *q* axis currents of the stator and $L_{f,i}$, $L_{af,i}$, $R_{f,i}$ are the self inductance of the field winding, mutual inductance between field and stator windings and the resistance of the field winding. The stator flux and voltage equations also written as

$$\begin{bmatrix} \psi_{d,i} \\ \psi_{q,i} \end{bmatrix} = -L_{s,i}(\delta_i) \begin{bmatrix} i_{d,i} \\ i_{q,i} \end{bmatrix} + L_{af,i} \begin{bmatrix} \cos(\delta_i) \\ \sin(\delta_i) \end{bmatrix} i_{f,i} \tag{3.1}$$

$$L_{s,i}(\delta_i) = L_{a0,i} I_{2\times 2} + \frac{3}{2} L_{a2,i} \begin{bmatrix} \cos(2\delta_i) & \sin(2\delta_i) \\ \sin(2\delta_i) & -\cos(2\delta_i) \end{bmatrix} \tag{3.2}$$

$$\begin{bmatrix} e_{d,i} \\ e_{q,i} \end{bmatrix} = \begin{bmatrix} \dot{\psi}_{d,i} \\ \dot{\psi}_{q,i} \end{bmatrix} + \omega_0 \begin{bmatrix} 0 & -1 \\ 1 & 0 \end{bmatrix} \begin{bmatrix} \psi_{d,i} \\ \psi_{q,i} \end{bmatrix} - R_{a,i} I_{2\times 2} \begin{bmatrix} i_{d,i} \\ i_{q,i} \end{bmatrix} \tag{4}$$

The electrical torque $T_{e,i}$ in (1.2) is calculated from

$$T_{e,i} = \frac{3}{2} p_f (\psi_{d,i} i_{q,i} - \psi_{q,i} i_{d,i}) \tag{5}$$

In the frequency range of the oscillations, the generator dynamics are governed by the mechanical swing (1) and the field winding (2) equations. It means the stator dynamics can be effectively ignored by setting $\dot{\psi}_{d,i} = \dot{\psi}_{q,i} = 0$. Hence, a choice of the state vector for the *i*th generator is

$$x_i = [\delta_i \quad \omega_i \quad \psi_{f,i}]^T \tag{6}$$

*B. The network*

We model the interconnection of the transmission lines, transformers, loads by a linear AC electric network with *m* ports that are connected to the *m* generators. For this purpose, it may be required to linearize some network components (such as the loads). Since the frequency of swing oscillation is much smaller than the synchronous speed $\omega_0$, we consider the network as a static subsystem (loads are assumed to be static in the frequency range of oscillations) to apply the steady state sinusoidal analysis. For a balanced three phase network, the result of modeling is an admittance matrix with *m* ports. The inputs to the network are the *m* voltage phasors for the ports and the outputs are the corresponding current phasors of the ports. The real part of the voltage phasor for the *i*th port is the element of *i*th stator voltage which is in the direction of the synchronous rotating frame which is $e_{d,i}$ and the imaginary part of the voltage phasor is the perpendicular element which is $e_{q,i}$. Hence, the voltage phasor of the *i*th port is $e_{d,i} + j e_{q,i}$ where *j* is the imaginary unit. Similarly, the current phasor for the *i*th port is $i_{d,i} + j i_{q,i}$. We also consider *m* disturbance currents in the network with pahsors $i_{d,i}^w + j i_{q,i}^w$. Based on these facts we can write

$$\begin{bmatrix} i_{d,1} \\ i_{d,2} \\ \vdots \\ i_{d,m} \end{bmatrix} + j \begin{bmatrix} i_{q,1} \\ i_{q,2} \\ \vdots \\ i_{q,m} \end{bmatrix} = Y(j\omega_0)\left(\begin{bmatrix} e_{d,1} \\ e_{d,2} \\ \vdots \\ e_{d,m} \end{bmatrix} + j \begin{bmatrix} e_{q,1} \\ e_{q,2} \\ \vdots \\ e_{q,m} \end{bmatrix}\right) + H(j\omega_0)\left(\begin{bmatrix} i^w_{d,1} \\ i^w_{d,2} \\ \vdots \\ i^w_{d,m} \end{bmatrix} + j \begin{bmatrix} i^w_{q,1} \\ i^w_{q,2} \\ \vdots \\ i^w_{q,m} \end{bmatrix}\right) \tag{7}$$

where $Y(j\omega_0)$ and $H(j\omega_0)$ are respectively the admittance matrix of the network and the disturbance transfer matrix at the synchronous frequency $\omega_0$.

### C. Composition

By regarding $u_i = e_{f,i}$, $w_i = [i^w_{d,i} \; i^w_{q,i}]^T$ respectively as the control and disturbance inputs to the $i$th generator, the set of equations (1) through (7) are a complete set of equations that can be solved for $\dot{x}_i$, $\psi_{d,i}$, $\psi_{q,i}$, $e_{d,i}$, $e_{q,i}$, $i_{d,i}$, $i_{q,i}$, $i_{f,i}$, $T_{e,i}$ ($1 \leq i \leq m$) in terms of $x_i$, $u_i$, $w_i$, $T_{m,i}$ ($1 \leq i \leq m$). By linearizing the solution around a working point we arrive at the small signal model of the power grid as

$$\dot{x} = Ax + B_u u + B_w w \tag{8.1}$$
$$x = [x_1^T \; x_2^T \; \cdots \; x_m^T]^T \tag{8.2}$$
$$u = [u_1^T \; u_2^T \; \cdots \; u_m^T]^T \tag{8.3}$$
$$w = [w_1^T \; w_2^T \; \cdots \; w_m^T]^T \tag{8.4}$$

## III. ANALYSIS TOOLS

Before proceeding to the main part of this work, in this section we present our time-discretization framework and the required tools for state feedback design to optimize either a quadratic cost function or an $H_\infty$ norm.

### A. Time-discretization with delay

In this part, time-discretization of a system with state $x \in \mathbb{R}^{n_x}$, control input $u \in \mathbb{R}^{n_u}$, disturbance input $w \in \mathbb{R}^{n_w}$, output $y \in \mathbb{R}^{n_y}$, and constant delay $d$ such that

$$\dot{x} = A_1 x + B_{1u} \bar{u} + B_{1w} w, \tag{9.1}$$
$$y = C_1 x + D_{1u} \bar{u} + D_{1w} w, \tag{9.2}$$
$$\hat{u}(t) = u(t - d) \tag{9.3}$$

together with a quadratic cost function

$$J = \int_0^\infty \begin{bmatrix} x(s) \\ \bar{u}(s) \end{bmatrix}^T \begin{bmatrix} Q_1 & N_1 \\ N_1^T & R_1 \end{bmatrix} \begin{bmatrix} x(s) \\ \bar{u}(s) \end{bmatrix} ds \tag{10}$$

is considered. Matrix $Q_1$ is constant and belongs to $\mathbb{R}^{n_x \times n_x}$. The other matrices are also constant with the appropriate dimensions. The cost function in (10) is computed for the case in which $w(t) = 0$ for every $t \in \mathbb{R}_0^+$. The dependence of $J$ on $\hat{u}$ (instead of $u$) will allow for a more sensible comparison of $J$ for different values of $d$.

The sampling instants are $kh$, $k \in \mathbb{Z}_0^+$. For every $k$, the state and output are sampled as $x_k = x(kh)$, $y_k = y(kh)$ and it is assumed that $u$ and $w$ are constant between the sampling instants such that $u(t) = u_k$ and $w(t) = w_k$ for $t \in (kh, kh+h]$.

The discretization involves the computation of output $y_k$ and cost function $J$ in terms of the discrete-time signals and a discrete state $z_k \in \mathbb{R}^{n_z}$ as

$$z_{k+1} = A_2 z_k + B_{2u} u_k + B_{2w} w_k \tag{11.1}$$

$$y_k = C_2 z_k + D_{2u} u_k + D_{2w} w_k \tag{11.2}$$

$$J = \sum_{i=0}^{\infty} \begin{bmatrix} z_i \\ u_i \end{bmatrix}^T \begin{bmatrix} Q_2 & N_2 \\ N_2^T & R_2 \end{bmatrix} \begin{bmatrix} z_i \\ u_i \end{bmatrix}. \tag{11.3}$$

The discrete state $z_k$ contains $x_k$ and a limited number of memory elements to capture the delay $d$ in (9.3).

The discretization is based on the following lemma

**Lemma 1**: if $\dot{x} = A_1 x + B_{1u} \bar{u}$ with $\bar{u}(t) = \hat{u}$ being constant for $t \in [a, a+b]$, then the following equations hold

$$x(a+b) = \Phi(b) x(a) + \Gamma(b) B_{1u} \hat{u} \tag{12.1}$$

$$\int_a^{a+b} \begin{bmatrix} x(s) \\ \hat{u} \end{bmatrix}^T \begin{bmatrix} Q_1 & N_1 \\ N_1^T & R_1 \end{bmatrix} \begin{bmatrix} x(s) \\ \hat{u} \end{bmatrix} ds = \begin{bmatrix} x(a) \\ \hat{u} \end{bmatrix}^T \Psi(b) \begin{bmatrix} x(a) \\ \hat{u} \end{bmatrix} \tag{12.2}$$

where

$$\Phi(\alpha) = \exp\{\alpha A_1\} \tag{13.1}$$

$$\Gamma(\alpha) = \int_0^\alpha e^{A_1 s} ds \tag{13.2}$$

$$\Psi(b) = \begin{bmatrix} \Psi_1(b) & \Psi_3(b) \\ \Psi_3^T(b) & \Psi_2(b) \end{bmatrix} \tag{14.1}$$

$$\Psi_1(b) = \Phi^T(b) P \Phi(b) - P \tag{14.2}$$

$$\Psi_3(b) = \Phi^T(b) P \Gamma(b) B_{1u} + \Phi^T(b) M - M \tag{14.3}$$

$$\Psi_2(b) = B_{1u}^T \Gamma^T(b) P \Gamma(b) B_{1u} + M^T \Gamma(b) B_{1u} + B_{1u}^T \Gamma^T(b) M + bU \tag{14.4}$$

$$Q_1 = PA_1 + A_1^T P, \tag{15.1}$$

$$N_1 = A_1^T M + P B_{1u}, \tag{15.2}$$

$$R_1 = B_{1u}^T M + M^T B_{1u} + U. \tag{15.3}$$

To save space, it is just mentioned that (12.1) is proved by replacement in the differential equation and (12.2) is proved by using (15) to show that

$$\frac{d}{ds} \begin{bmatrix} x(s) \\ \hat{u} \end{bmatrix}^T \begin{bmatrix} P & M \\ M^T & Us \end{bmatrix} \begin{bmatrix} x(s) \\ \hat{u} \end{bmatrix} = \begin{bmatrix} x(s) \\ \hat{u} \end{bmatrix}^T \begin{bmatrix} Q & N \\ N^T & R \end{bmatrix} \begin{bmatrix} x(s) \\ \hat{u} \end{bmatrix}.$$

By applying Lemma 1 for the case of $d = 0$ we have

$$\begin{aligned} A_2 &= \Phi(h), & C_2 &= C_1, \\ B_{2u} &= \Gamma(h) B_{1u}, & D_{2u} &= D_{1u}, \\ B_{2w} &= \Gamma(h) B_{1w}, & D_{2w} &= D_{1w}. \end{aligned} \tag{16.1}$$

$$\begin{bmatrix} Q_2 & N_2 \\ N_2^T & R_2 \end{bmatrix} = \begin{bmatrix} \Psi_1(h) & \Psi_3(h) \\ \Psi_3^T(h) & \Psi_2(h) \end{bmatrix} \tag{16.2}$$

$$z_k = x_k \tag{16.3}$$

with $\Psi_1, \Psi_2, \Psi_3$ given in (14), (15) and $z_k$ being equal to $x_k$.

If $d > 0$, then we write $d = qh + r$ such that

$$q = \max\{k \in \mathbb{Z}: qh < d\} \quad (17.1)$$
$$r = d - qh \quad (17.2)$$

To applying Lemma 1 when $d > 0$ we consider that for $t \in (kh, kh+r]$ we have $\bar{u}(t) = u_{k-q-1}$ and for $t \in (kh+r, kh+h]$ we have $\bar{u}(t) = u_{k-q}$ and we should apply the lemma successively twice. The result can be compactly presented as

$$\begin{bmatrix} A_2 & B_{2u} \\ C_2 & D_{2u} \end{bmatrix} = \begin{bmatrix} \Phi(h) & \Gamma_1(r) & \Gamma_0(r) & 0 \\ 0 & 0 & I & 0 \\ 0 & 0 & 0 & I_{qn_u} \\ C_1 & 0 & D_{2u} & 0 \end{bmatrix} \quad (18.1)$$

$$\Gamma_1(r) = \Phi(h-r)\Gamma(r)B_{1u} \quad (18.2)$$
$$\Gamma_0(r) = \Gamma(h-r)B_{1u} \quad (18.3)$$
$$B_{2w} = \begin{bmatrix} \Gamma_w^T & 0_{(q+1)n_u \times n_w}^T \end{bmatrix}^T, \quad D_{2w} = D_{1w} \quad (18.4)$$

$$\begin{bmatrix} Q_2 & N_2 \\ N_2^T & R_2 \end{bmatrix} = \text{diag}\left\{ \begin{bmatrix} \Psi(r) & 0 \\ 0 & 0 \end{bmatrix} + \right.$$
$$\left. \Phi_1^T(r)\Psi(h-r)\Phi_1(r), 0_{(q-1)h} \right\} \quad (18.5)$$

$$\Phi_1(\alpha) = \begin{bmatrix} \Phi(\alpha) & \Gamma(\alpha)B_{1u} & 0 \\ 0 & 0 & I \end{bmatrix} \quad (18.6)$$

$$z_k = \begin{bmatrix} x_k^T & u_{k-q-1}^T & u_{k-q}^T & \cdots & u_{k-1}^T \end{bmatrix}^T, \quad (18.7)$$

If $q$ is zero ($d \leq h$), then the fourth row and column on the right side of (18.1) become null (eliminated) and $B_{2u}$, $D_{2u}$ are obtained from the third column (otherwise, they are obtained from the forth column).

*B. Linear quadratic regulator*

For the system (11.1) with $w_k = 0$, $k \in \mathbb{Z}_0^+$ and the cost function $J$ in (11.3), if the pair $(A_2, B_{2u})$ is controllable (according to [20]) then the optimal controller that minimizes $J$ is a state feedback gain $F_{LQR}$ given by

$$F_{LQR} = -(B_{2u}^T P B_{2u} + R_2)^{-1}(B_{u2}^T P A_2 + N_2^T) \quad (19)$$

where $P$ is calculated from the following discrete-time algebraic Riccati equation

$$A_2^T P A_2 - P - (A_2^T P B_{2u} + N_2)(B_{2u}^T P B_{2u} + R_2)^{-1} \times$$
$$(B_{u2}^T P A_2 + N_2^T) + Q_2 = 0 \quad (20)$$

and the optimal value of $J$ is given by

$$J^* = z_0^T P z_0. \quad (21)$$

*C. $H_\infty$ state feedback design*

For the system (11.1), (11.2), if the conditions in [21, Theorem 9.4] are satisfied, then by applying the state feedback

$$F_{dist} = H_1^{-1}(B_{2u}^T P A_2 + D_{2u}^T C_2 + H_2 H_3^{-1} H_4) \quad (22)$$

with the algebraic Riccati equation for $P$ and definitions of $H_1$ through $H_6$ given as

$$P = A_2^T P A_2 + C_2^T C_2 - H_5^T H_6^{-1} H_5 \tag{23.1}$$

$$H_1 = B_{2u}^T P B_{2u} + D_{2u}^T D_{u2} + H_2 H_3^{-1} H_2^T \tag{23.2}$$

$$H_2 = B_{2u}^T P B_{2w} + D_{2u}^T D_{2w} \tag{23.3}$$

$$H_3 = \gamma^2 I - D_{2w}^T D_{2w} - B_{2w}^T P B_{2w} \tag{23.4}$$

$$H_4 = B_{2w}^T P A_2 + D_{2w}^T C_2 \tag{23.5}$$

$$H_5 = B_2^T P A_2 + D_2^T C_2 \tag{23.6}$$

$$H_6 = D_2^T D_2 - \text{diag}\{0, \gamma^2 I_{n_w}\} + B_2^T P B_2 \tag{23.7}$$

$$B_2 = [B_{2u}\ B_{2w}],\ D_2 = [D_{2u}\ D_{2w}] \tag{23.8}$$

for some $\gamma \in \mathbb{R}^+$, then it is guaranteed that the $H_\infty$ norm from $w_k$ to $y_k$ defined as $G_{yw} = \sup_w (\sum_{k=0}^{\infty} y_k^2)^{½} (\sum_{k=0}^{\infty} w_k^2)^{-½}$ is less than $\gamma$.

IV. DISTRIBUTED NETWORKED CONTROL SCHEME

Consider a large plant (e.g. the power system) which is described by (8.1) with state and input decompositions in (8.2) through (8.4). The objective of this section is to design a distributed networked control system (DNCS) for such a plant assuming that there is a transmission delay $d_{ik}$ in the link from the $k$th subsystem to the $i$th subsystem.

We consider two different performance measures for the control design problem. The first one is a cost function

$$J = \int_0^\infty \left( x^T(s) Q x(s) + u(s) R u(s) \right) ds \tag{25}$$

and the second one is the $H_\infty$ norm from $w$ to $y$ with

$$y = Cx + D_u u + D_w w \tag{26}$$

as a disturbance attenuation criterion.

The delays are assumed to satisfy two conditions:

- The delays are constant since there usually exist a dedicated and reliable communication infrastructure with constant traffic for automating a critical system which eliminates the randomness of communication.

- The delays are symmetric i.e. $d_{ik} = d_{ki}$ for every $1 \leq i, k \leq m$ due the fact that almost all communication link are full duplex (bidirectional).

Assuming that each subsystem has access to its own state information, the DNCS is composed of local control signals $K_i x_i$, and remote control signals $\bar{u}_i$, $1 \leq i \leq m$ such that

$$u_i = K_i x_i + \bar{u}_i \tag{27.1}$$

$$\bar{u}_i(t) = v_{i,k}\ \text{for}\ t \in (kh+d_i,\ kh+h+d_i] \tag{27.2}$$

The local control signal $K_i x_i$ is updated continuously and the remote control signal $\bar{u}_i$ is updated at instants $kh+d_i$, $k \in \mathbb{Z}_0^+$ and remains equal to $v_{i,k}$ until $kh+h+d_i$ according to (27.2). In fact, the $i$th subsystem waits during $[kh, kh+d_i)$ such that all of the information required for calculation of $v_{i,k}$ is received. The values of $d_i$ depends on $d_{ik}$ for $1 \leq i, k \leq m$ and will be determined in the remaining.

In this work, we focus on the design of the remote control signal $\bar{u}_i$, $1 \leq i \leq m$ by assuming that the local gains $K_i$, $1 \leq i \leq m$, are already designed and known.

By combining (8.1) and (27.1) we have

$$\dot{x} = \bar{A}x + B_u\bar{u} + B_w w \tag{28.1}$$

$$\bar{A} = A + B_u K \tag{28.2}$$

$$\bar{u} = [\bar{u}_1^T \ \bar{u}_2^T \ \cdots \ \bar{u}_m^T]^T \tag{28.3}$$

$$K = \text{diag}\{K_1, K_2, \cdots, K_m\} \tag{28.4}$$

We apply the changes of coordinate

$$x = M_x \hat{x}, \quad \hat{x} = [\hat{x}_1^T \ \hat{x}_2^T \ \cdots \ \hat{x}_{\hat{m}}^T]^T \tag{29.1}$$

$$\bar{u} = M_u \hat{u}, \quad \hat{u} = [\hat{u}_1^T \ \hat{u}_2^T \ \cdots \ \hat{u}_{\hat{m}}^T]^T \tag{29.2}$$

$$w = M_w \hat{w}, \quad \hat{w} = [\hat{w}_1^T \ \hat{w}_2^T \ \cdots \ \hat{w}_{\hat{m}}^T]^T \tag{29.3}$$

such that

$$M_x^{-1} \bar{A} M_x = \text{diag}\{\hat{A}_1, \hat{A}_2, \cdots, \hat{A}_{\hat{m}}\}, \tag{30.1}$$

$$B_u M_u = \text{diag}\{\hat{B}_1^u, \hat{B}_2^u, \cdots, \hat{B}_{\hat{m}}^u\}, \tag{30.2}$$

$$B_w M_w = \text{diag}\{\hat{B}_1^w, \hat{B}_2^w, \cdots, \hat{B}_{\hat{m}}^w\} \tag{30.3}$$

with appropriate dimensions for the new matrices and variables such that we can decompose (28.1) as

$$\dot{\hat{x}}_i = \hat{A}_i \hat{x}_i + \hat{B}_i^u \hat{u}_i + \hat{B}_i^w \hat{w}_i, \quad 1 \le i \le \hat{m}. \tag{31}$$

**Remark 2**: In general, the problem of finding $M_x$ has many solutions (including the Jordan canonical form). It would be desirable to find $M_x$ such that the new variables $\hat{x}_i$, $1 \le i \le m$, have meaningful interpretations. Ones $M_x$ is determined, the problems of finding $M_u$ and $M_w$ are similar to Gaussian elimination that are solvable in general if $\hat{m}$ is larger than the dimension of $\hat{x}_i$ for every $1 \le i \le m$. Otherwise, $M_u$ and $M_w$ may exist in special cases (such as the case in the next section). We ignore further discussion on finding the new coordinates due to the space limits and focus on a symmetrical dual machine power grid in the next section.

In the new coordinates, the cost function $J$ in (25) is written as (32) and the output in (26) is written as (33).

$$J = \int_0^\infty \begin{bmatrix} \hat{x}(s) \\ \hat{u}(s) \end{bmatrix}^T \hat{U} \begin{bmatrix} \hat{x}(s) \\ \hat{u}(s) \end{bmatrix} ds \tag{32}$$

$$\hat{U} = \begin{bmatrix} M_x & 0 \\ 0 & M_u \end{bmatrix}^T \begin{bmatrix} Q + K^T R K & K^T R \\ R K & R \end{bmatrix} \begin{bmatrix} M_x & 0 \\ 0 & M_u \end{bmatrix}$$

$$y = \hat{C}\hat{x} + \hat{D}_u \hat{u} + \hat{D}_w \hat{w} \tag{33}$$

$$\hat{C} = C M_x, \quad \hat{D}_u = D_u M_u, \quad \hat{D}_w = D_w M_w$$

Since the control structure in (27) is not an entire state feedback (from $x$), it is not possible to find the gains in order to optimize the selected performance measures. Because, the optimal gains will depend on the initial conditions of the system in general. However, we can design optimal state feedbacks for each of the subsystems in (31). In fact when we design the control for the $i$th subsystem, we assume that $\hat{x}_k = 0$ for every $k \ne i$. Stability of each subsystem (due to the optimal control) will ensure the stability of the entire system. In this way the cost function $J$ in (32) and the output $y$ in (33) are reduced to $J_i$ in (34) and $y_i$ in (35) for the $i$th subsystem.

$$J_i = \int_0^\infty \begin{bmatrix}\hat{x}_i(s)\\\hat{u}_i(s)\end{bmatrix}^T \hat{U}_i \begin{bmatrix}\hat{x}_i(s)\\\hat{u}_i(s)\end{bmatrix} ds \tag{34}$$

$$\hat{U}_i = \text{diag}\{E_i^x, E_i^u\}^T \hat{U} \text{diag}\{E_i^x, E_i^u\}$$

$$y_i = \hat{C}_i \hat{x}_i + \hat{D}_i^u \hat{u}_i + \hat{D}_i^w \hat{w}_i \tag{35}$$

$$\hat{C}_i = CE_i^x, \quad \hat{D}_i^u = \hat{D}_u E_i^u, \quad \hat{D}_i^w = \hat{D}_w E_i^w$$

where the matrices $E_i^x, E_i^u$ and $E_i^w$ are such that $\hat{x} = \sum_{i=1}^{\hat{m}} E_i^x \hat{x}_i$, $\hat{u} = \sum_{i=1}^{\hat{m}} E_i^u \hat{u}_i$, $\hat{w} = \sum_{i=1}^{\hat{m}} E_i^w \hat{w}_i$.

The delays of $\bar{u}$ in (27.2) are reflected in $\hat{u}$ according to (29.2). We consider the delays in $\hat{u}$ in the form of

$$\hat{u}_i(t) = \hat{v}_{i,k} \text{ for } t \in (kh + \hat{d}_i, kh + h + \hat{d}_i]. \tag{36}$$

Based on the above equation we discretize (31), (34) and (35) with delays $\hat{d}_i$, $1 \le i \le \hat{m}$ using the procedure of part $A$ in section III. The result is written as

$$z_{i,k+1} = A_i^z z_{i,k} + B_i^v \hat{v}_{i,k} + B_i^w \hat{w}_{i,k} \tag{37.1}$$

$$y_{i,k} = C_i^z z_{i,k} + D_i^v \hat{v}_{i,k} + D_i^w \hat{w}_{i,k} \tag{37.2}$$

$$J_i = \sum_{k=0}^{\infty} \begin{bmatrix}z_{i,k}\\u_{i,k}\end{bmatrix}^T U_d \begin{bmatrix}z_{i,k}\\u_{i,k}\end{bmatrix}. \tag{37.3}$$

By applying either of the control design methods in parts III.B and III.C we obtain control laws in the following form

$$\hat{v}_{i,k} = F_i z_{i,k}, \quad 1 \le i \le \hat{m} \tag{38}$$

To find the relationship between $v_\rho, d_\rho, 1 \le \rho \le m$ and $\hat{v}_i, \hat{d}_i, 1 \le i \le \hat{m}$, we write (29.2) as

$$\bar{u}_\rho = \sum_{i=1}^{\hat{m}} [M_u]_{\rho i} \hat{u}_i \tag{39}$$

with $[M_u]_{\rho i}$ being the appropriate block of $M_u$. Since $\bar{u}_\rho$ and $\hat{u}_i$ are the delayed versions of $v_\rho$ and $\hat{v}_i$ we must have

$$v_{\rho,k} = \sum_{i=1}^{\hat{m}} [M_u]_{\rho i} \hat{v}_{i,k}. \tag{40}$$

$$[M_u]_{\rho i} \ne 0 \Rightarrow d_\rho = \hat{d}_i \quad \forall \ 1 \le \rho \le m, 1 \le i \le \hat{m}. \tag{41}$$

in order to obtain $\bar{u}$ from $\hat{u}$ as in (39).

Based on (41) we obtain $d_\rho$ as below.

$$d_\rho = \max_i \{\hat{d}_i : [M_u]_{\rho i} \ne 0\} \tag{42}$$

Due to the distributed nature of the system, the controller for the $\alpha$th subsystem must calculate (38) for every $i$ with $[M_u]_{\alpha i} \ne 0$. This requires $\hat{x}_i(kh)$ for every $[M_u]_{\alpha i} \ne 0$. From (29.1) we have $\hat{x}_i(kh) = \sum_{\beta=1}^m [M_x^{-1}]_{i\beta} x_\beta(kh)$. Hence, to obtain $\hat{x}_i(kh)$ the controller must wait until $x_\beta(kh)$ is received for every $[M_x^{-1}]_{i\beta} \ne 0$ and we can write

$$\hat{d}_i \ge \max_\alpha \{\max_\beta \{d_{\alpha\beta} : [M_x^{-1}]_{i\beta} \ne 0\} : [M_u]_{\alpha i} \ne 0\}$$

By considering the minimum time required for information gathering we can write

$$\hat{d}_i = \max_{\alpha,\beta} \{d_{\alpha\beta} : [M_u]_{\alpha i} \ne 0, [M_x^{-1}]_{i\beta} \ne 0\}. \tag{43}$$

## V. THE SYMMETRICAL DUAL-MACHINE POWER SYSTEM

In this section we apply the results of the previous sections to a symmetrical dual machine power system, i.e. $m$ in (8) is 2 and the generators are identical. The generator parameters are $L_{a0}$ = 4.9mH, $L_{a2}$ = 46µH, $L_f$ = 577mH, $L_{af}$ = 4 mH, $R_a$ = 3 m$\Omega$, $R_f$ = 71.5 m$\Omega$, $J$ = 27548 Kg.m$^2$, $p_f$ = 2, $B$ = 10 Kg.m$^2$/s and $w_0$ = 377 rad/s. Each generator is connected to a load with impedance $Z_L$ = 6.2+$j$2.1 via a transmission line with impedance $Z_T$ = 0.011+$j$0.106. The load terminals are connected via a transmission line with impedance $Z_C$ = 0.054+$j$0.53. At each load terminal we apply a current disturbance input with two elements for the active and reactive components. The cost function (25) is selected as $J = \int_0^\infty [\delta_1^2 + \delta_2^2 + 2.5\times10^{-5}(e_{f,1}^2 + e_{f,2}^2)]ds$ and the output in (26) is selected as $y = [\delta_1\ \delta_2]^T + 10^{-2}[e_{f,1}\ e_{f,2}]^T$ by indicating on the swing modes $\delta_1$ and $\delta_2$. The local control gains for the case of cost function based design are selected as $K_1$ = [169.6  201  −3.04] and for the case of disturbance attenuation based design as $K_2$ = [544750  700010  -9890] by local designs.

In the selected symmetrical grid the oscillations are caused by a difference between the variables of the two generators. Hence, we define an oscillation mode $\hat{x}_1 = x_1 - x_2$ and a common mode $\hat{x}_2 = x_1 + x_2$, ($x_i$ is the state of the $i$th generator). The changes of coordinates for the inputs are also obtained as $\hat{u}_1 = u_1 - u_2$, $\hat{w}_1 = w_1 - w_2$, $\hat{u}_2 = u_1 + u_2$, $\hat{w}_2 = w_1 + w_2$. By applying the design procedure of the previous section, the cost function $J$ and the attenuation from the current disturbance (as the performance measures) for the oscillation mode (which captures the oscillations) are plotted in Figures 1 and 2 respectively. The value of each performance measure for the case of decentralized control (only local gains $K_1$ and $K_2$) as an upper bound, and global state feedback (simultaneous design of local and remote controls without delays) as a lower bound of the measures are also plotted in dotted format. The plots show the ability of the DNCS to dampen the swing oscillations for different values of the delay from two different viewpoints. Due to the variable costs of the communication infrastructures with different latencies and reliability, the provided analysis is useful for deciding on the choice of the communication technology for the DNCS as well as the DNCS design.

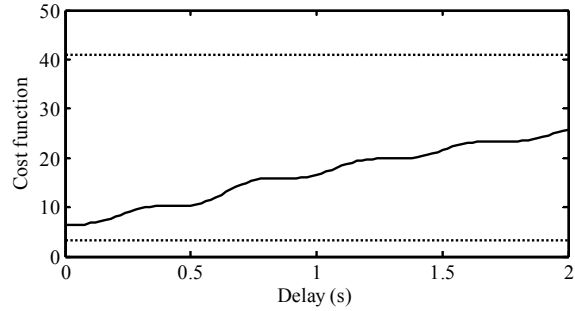

Figure 1.  Optimal oscillation mode cost function vs. the delay

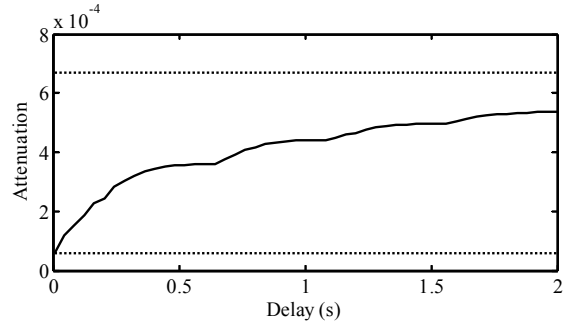

Figure 2.  Optimal oscillation mode disturbance attenuation vs. the delay

## VI. Conclusion

A framework was presented for distributed networked control system (DNCS) design for large systems based on the linear quadratic cost and $H_\infty$ disturbance attenuation performance measures. Since a global optimal design is not possible, the design is based on optimality of different modes. The results were applied to a symmetrical dual machine power grid to evaluate the ability of the DNCS in dampening of the swing oscillations.